# OntoLoki: an automatic, instance-based method for the evaluation of biological ontologies on the Semantic Web[1]

**Benjamin M. Good, Gavin Ha, Chi K. Ho, Mark D. Wilkinson**

## Background

In recent years, the explosive growth of data in the life sciences has produced the need for many new biological and medical ontologies [1, 2]. To meet this need, the biomedical informatics community has developed powerful tools for building ontologies such as Protégé [3, 4], established institutions devoted to improving and expanding the practice of biomedical ontology design [5], and produced an ever-increasing number of ontologies and ontology-based applications [6]. Outside of the biomedical community, the broader W3C-lead semantic Web initiative has produced new, standardized languages for sharing ontologies on the Web, such as OWL [7], as well as efficient algorithms for automated, deductive reasoning over the knowledge represented within them [8, 9]. These advances aid the processes of building, sharing, and using biological and medical ontologies; however, one well-recognized, yet largely unmet need remains the development of effective, objective methods for the evaluation of ontology quality [10, 11]. As Jeremy Rogers points out,

*"the medical and non-medical ontology engineering communities have yet to define, much less to regularly practice, a comprehensive and systematic methodology for assuring, or improving, the quality of their product"* [12].

In much the same way that consistent standards for designing, building, and evaluating the products of physical engineering efforts can contribute to the efficiency of the engineering process and the reliability of the end-product, an effective, systematic methodology for ontology design and evaluation would be a great benefit to the community. Though some attempts have been made at comprehensive methodologies, such as Methontology [13], none has gained widespread adoption. We suggest that before such a methodology can truly be successful, a wealth of measurements of the characteristics of ontologies as well as a detailed understanding of their implications is necessary.

To contribute to the eventual development of a comprehensive methodology for ontology design and evaluation, we introduce a new way of automatically measuring one important ontology characteristic that is currently inaccessible using existing methods. The delineation of precise definitions for each class in an ontology and the consistent application of these definitions to the assignment of instances to classes are well-accepted desiderata for ontologies. If ontologies are defined with formal restrictions on class membership, then such consistency can be checked automatically using existing technology. If no such logical restrictions are applied however, as is the case with many current biological and medical ontologies, then there are currently no automated methods for measuring the consistency of instance assignment for the purpose of evaluation. The aim of this study is thus to identify, implement, and test a new method for automatic, data-driven ontology evaluation that is suitable for the evaluation of the consistency of ontologies with no formally defined restrictions on class membership.

---

[1] This manuscript corresponds to Chapter 4 in Benjamin Good's Ph.D. Dissertation "Strategies for amassing, characterizing, and applying third-party metadata in bioinformatics". (2009) University of British Columbia, Department of Bioinformatics. http://hdl.handle.net/2429/7115

**Results**

The results of this research comprise the OntoLoki method for ontology evaluation, its implementation, and the empirical evaluation of the implementation. Each aspect is now presented in turn.

**OntoLoki method**

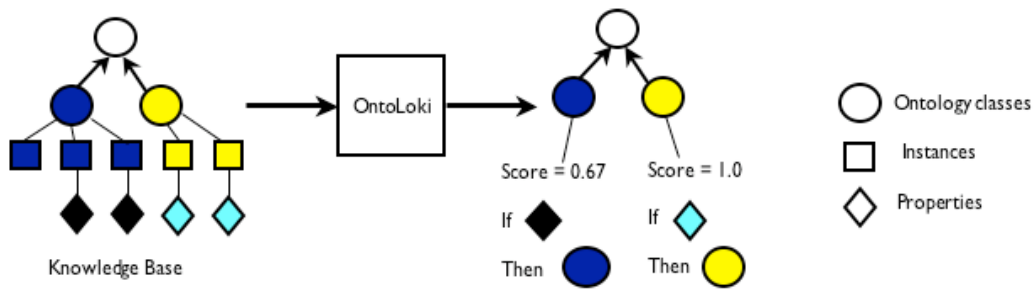

**Figure 1. The input and the output for the OntoLoki method.** The properties assigned to each instance are used to learn classification rules whose performance is quantified. The rule learned for the dark blue class correctly classifies two of its three instances based on the property indicated by the black diamonds. The rule learned for the yellow class correctly predicts all of its instances based on the presence of the property indicated by the light blue diamonds.

**Input and output**

Figure 1 illustrates the basic process of ontology evaluation using the OntoLoki method. The input is a knowledge base containing an ontology, instances assigned to the classes in that ontology, and properties associated with those instances. The output is, for each class in the ontology,

1. a rule that defines a pattern of properties that attempts to separate the instances of the class from the instances of other classes,
2. a quantitative score, called 'classification consistency', based on the ability of the identified rule to correctly predict class membership for the instances in the knowledge base.

High classification consistency means that instances are assigned to classes based on strictly followed rules expressed through specific patterns of properties evident in the instances. For example, a classification rule could be used to assign all instances in a particular knowledge base that displayed both the property of having blue feathers and the property of having a red beak to a particular class. As long as the rule is applied reliably, such that all members of the class exhibit this pattern of properties, the class should receive a high classification consistency score. On the other hand, a low classification consistency score is an indication that – in the context of the specific knowledge base used in the evaluation – no consistent pattern of properties could be found to differentiate the members of the class in question from the other classes.

**Processing**

The classification consistency of an ontology, effectively the ability to identify patterns of properties that can be used to correctly assign instances to its classes, may be computed as follows:

1. Assemble a knowledge base composed of the ontology, instances of each of the classes considered from the ontology, and the properties of interest for those instances.
2. For each class in the ontology,
    a. Compose a dataset consisting of instances of that class (positive examples) and instances that are not members of that class (negative examples). This data may take the form of a simple table, where the rows correspond to instances and the columns correspond to properties of those instances.
    b. Use this data to build a classifier to distinguish between positive and negative instances of the class in question.
        i. Generate scores for each class based on the ability of the classifier to correctly predict the classes associated with the instances based on their properties.
        ii. Record the pattern identified by the classifier for each class.

To provide a single classification consistency score for the entire ontology, report the average score across all the classes.

**Implementation**

The implementation presented here is designed for application to ontologies and knowledge bases created according to the standards defined by the semantic Web working group(s) of the World Wide Web Consortium. Ontologies are thus expected to be represented in the OWL language [7] and knowledge bases are assembled by building or discovering RDF [14] graphs containing instances of the ontology classes. The RDF predicates originating from each instance of an ontology class to be evaluated provide the properties of those instances used for the evaluation. Chains of such predicates and their objects can be combined to form arbitrarily complex properties associated with each instance.

**Step 1 – knowledge base discovery**

The first step in the process of conducting an OntoLoki evaluation is the discovery of a knowledge base containing instances of the ontology classes and their properties. Though we would eventually like to include a discovery module in our implementation that assembles the required knowledge bases dynamically by identifying suitable content on the semantic Web, for example, through the automatic generation of queries issued to semantic repositories such as Bio2RDF [15, 16] and DBpedia [17], the current implementation requires manual intervention during the composition of the needed datasets. Once an OWL/RDF knowledge base is assembled, the rest of the processing is completely automated. In the experiments presented below, all the data is gathered through queries to the UniProt protein database [18]. This resource was selected because the instances of the evaluated ontologies correspond to proteins and the queries to this resource return RDF representations of the properties associated with proteins. To clarify interpretation of the experimental results below and to illustrate the raw input to our implementation, we provide a short description of the UniProt data.

**UniProt beta RDF representation**

The RDF versions of UniProt protein records are organized according to the UniProt core ontology [19]. At the time of writing, this OWL ontology used 151 classes, 39 object properties, 65 data properties, and 59 individuals to form the descriptions of all of the UniProt proteins. Each protein resource is represented as an instance of the class 'Protein'. The two most important predicates associated with proteins for present purposes are 'annotation' and 'seeAlso'.

The UniProt 'annotation' object property is used to form relations between protein instances and instances of the class 'Annotation'. 'Annotation' is subdivided into many different kinds of annotation, including 83 distinct subclasses. Examples of Annotation classes include 'Transmembrane_Annotation', 'Helix_Annotation', 'Signal_Peptide_Annotation', 'Lipidation_Annotation', and 'Biotechnology_Annotation'. These classes are used in at least two importantly different ways. In one, the class is specific enough to thoroughly encapsulate the intended meaning of the annotation; in the other, the precise meaning must be inferred from additional, often non-formal, statements. For example, consider the sample of UniProt RDF statements about the protein P35829 depicted in Figure 2 [20].

In Figure 2, protein P35829 is described with three annotations. The first is a "Signal Peptide Annotation", the second a "Function Annotation", and the third a "Post-translational modification annotation". The Signal Peptide annotation is quite specific, needing no more additional clarification as to its meaning. On the other hand, the latter two of these annotations are ambiguous and are thus enhanced in the record with textual comments that contain most of their meaning (e.g. the Post-translational modification is 'Glycosylated').

This situation raises a problem for automated semantic processing in that it requires the implementation to either ignore the information present in the comment fields, or incorporate some manner of natural language understanding to extract the information. As the aim of the implementation generated for this project is to take advantage of the *structured* data of the semantic Web, the comments were not incorporated into the training data for the classifiers.

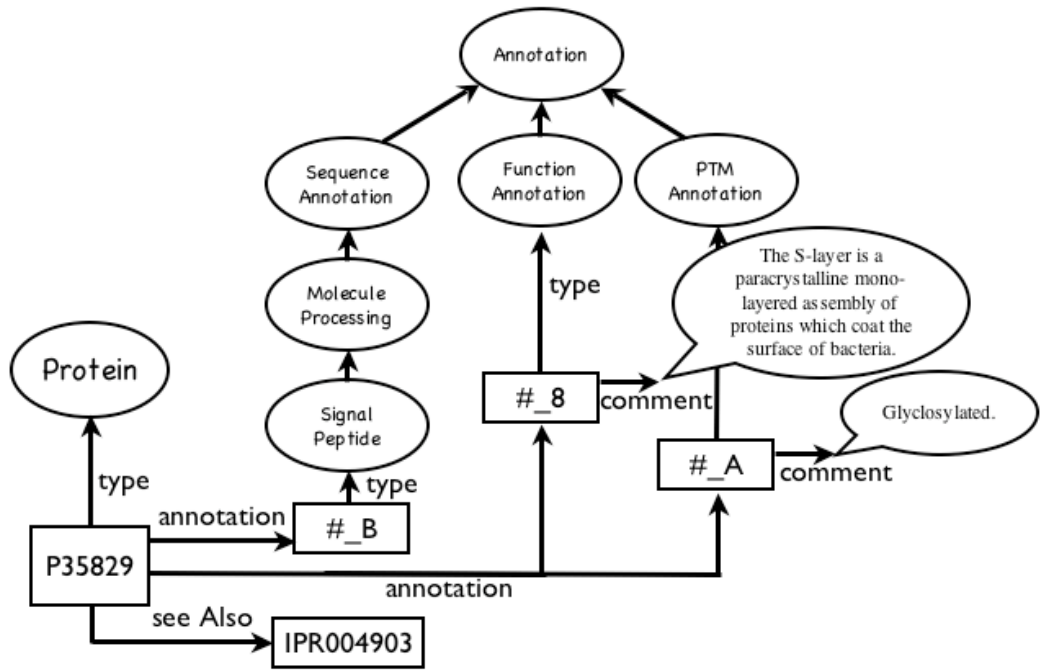

**Figure 2. Sample RDF description of the UniProt protein P35829, an S-layer protein precursor from Lactobacillus acidophilus.** Classes are indicated with ovals, instances with square boxes, predicates with arrows, and Strings as 'idea bubbles'. Some instances – for example, '#_B' – are created to fill the place of the blank nodes needed to create the descriptions of the instance under consideration ('P35829'). P35829 has three annotations: a signal peptide annotation, a function annotation, and a PTM annotation.

The other important property used to describe proteins in the UniProt data is rdfs:seeAlso. This is used to form relations between the proteins in UniProt and records from other databases. For example, P35829 is the subject of the following statement:

<rdfs:seeAlso rdf:resource="http://purl.uniprot.org/interpro/IPR004903"/>

Unfortunately, both the nature of these relationships and the nature of the external record is not captured. From the standpoint of an agent processing this data, each of the above statements is equally meaningless; each saying P35829 has some relationship to some other resource of unknown type/nature. Though it goes against the goal of providing a completely generic implementation that works according to the standards of the semantic Web, the information about the InterPro domains was needed for this particular experiment. Thus, a specific block of code was added to extract the InterPro annotations and add them to the training data. The other seeAlso statements were not processed.

### Step 2 – propositionalization of the RDF

Once an appropriate OWL/RDF knowledge base is assembled, our implementation maps the relational structure of the RDF graph to the flat structure of the tables used by most current class prediction algorithms in a process known as 'propositionalization' [21]. Each instance becomes a row in a simple 2-dimensional table and each property becomes a column. To identify properties, the RDF graph is traversed until either pre-defined endpoints or a specified maximum depth is reached. For example, properties that would be extracted from the protein graph represented in Figure 2 would be 'annotation_Signal-Peptide', 'annotation_Function- Annotation', and 'seeAlso-IPR004903'. More complex properties can be extracted by traversing deeper into the graph and recording the path.

The following list delineates important choices made in our implementation that may or may not be appropriate in others.

1. The extracted table represents a set of binary values. Each instance is assigned a true or a false value for each identified property.
2. The table does not reflect the open-world assumption of the semantic Web. If a property is not attached to an instance in the knowledge base, it is assumed not to exist and set to false.
3. The table does represent the subsumption relationship. Instances are indicated as members of both the class they are directly assigned to and all of the superclasses of that class.

**Step 3 – preparing the dataset for the evaluation of a single class**

Once a table representing the knowledge base is constructed, subsections of it are extracted for the evaluation of individual classes. To begin the evaluation of any particular class in the ontology, a table of data is composed specifically for that class by:

1. Selecting only those rows that are instances of the direct parent of the class (where simple subsumption reasoning ensures that this set will contain all of the instances of the class in question).
2. Removing all of the class columns (is a) except for the class in question.

This results in a table with one class column corresponding to the class being evaluated where all the rows correspond either to instances of that class (positive examples), or instances of its parent or sibling classes that are not instances of it (negative examples). This formulation allows for the detection of patterns that serve to discriminate classes from their parents and, through subsumption, their hierarchical siblings. If such patterns are detected, the addition of the class in question to that particular hierarchical location in the ontology is considered justified and the reason for its inclusion denudated.

**Step 4 – classifier training and testing**

Once a table containing positive and negative instances of a particular class is created, the next step is to find patterns of properties that can define the rules for membership in the class. In our implementation, this inductive process is accomplished through the application of supervised learning algorithms made available by the Waikato Environment for Knowledge Analysis (WEKA)[22].

Of the many learning algorithms available through WEKA, just the following were utilized in the experiments described below.

1. ZeroR simply always predicts the class displayed by the majority of the instances

in the training set and thus serves as a baseline for evaluating the other algorithms.
2. OneR creates a rule based on the most informative single attribute in the training data.
3. JRip is an implementation of the Ripper rule induction algorithm [23] capable of identifying fairly complex, but easily interpreted classification rules.
4. Chi25_JRip. In this case we use WEKA's AttributeSelectedClassifier meta–classifier. First, all of the attributes (properties of the instances) are ranked using the value of the chi-squared statistic for the attribute with respect to the class. Then, only the top 25 attributes are used to train and test the JRip algorithm.

These were selected as demonstrative cases because they are relatively fast and yield easily interpretable rules; however, any particular evaluation might benefit from a different classifier. With this in mind, our implementation makes it possible to select any of the hundreds learning algorithms provided by WEKA when running an evaluation.

**Quantifying classification consistency**

The classification rules identified by the learning algorithms need to be quantified based on their ability to separate the positive and negative instances for the class in question. Following a traditional approach to evaluating machine learning schemes with limited available data, the classifiers (learning algorithms) are repeatedly trained on ninety percent of the data and then tested on the remaining ten percent until all ten such divisions are complete [22]. For each such 10-fold cross-validation experiment, the numbers of all the correct and incorrect predictions made on the testing data are recorded. From these, various estimates of the quality of the class predictor (the decision model learned) are derived and reported to indicate the consistency of each class. The metrics reported by our implementation (as they are implemented in WEKA) are:

1. Accuracy = $N(C)/N(T)$ where $N(C)$ equals the number of correct predictions and $N(T)$ equals the total number of predictions.
2. F-Measure for predicting the target class $F(1)$ and for predicting not the target class $F(0)$ a. $F(c) = 2(P(c) *R(c))/P(c)+R(c)$ where $P(c)$ equals the precision of predicting the class c and $R(c)$ equals the recall for predicting class c.
3. Kappa: $K = (P(A) – P(E))/(1-P(E))$ where $P(A)$ equals the probability of correct prediction of the learned classifier and $P(E)$ equals the probability of correct prediction by chance.
4. Kononenko-Bratko information gain. See [24] for the derivation of this metric which is generally similar in intent to the Kappa statistic.

These estimates of the quality of the induced decision models attempt to reflect the relative existence of a consistent, unique pattern of properties associated with the class being evaluated.

**Testing**

Now that the method, illustrated in Figure 3, and our implementation have been described, we turn to an assessment of the data that it produces. We assert that the OntoLoki metrics quantitatively represent the quality for an ontology in a particular context and that the classification rules learned in the process could be useful in both the process of ontology engineering and in knowledge discovery. To provide evidence for these claims, we now present the results of the evaluation of several ontologies. The evaluations begin with a control experiment intended to show that the implementation is successful in producing the expected results on artificially generated ontologies displaying known levels of quality. This is followed by the evaluation of two real-world, biological ontologies, one very simple and one large and structurally complex.

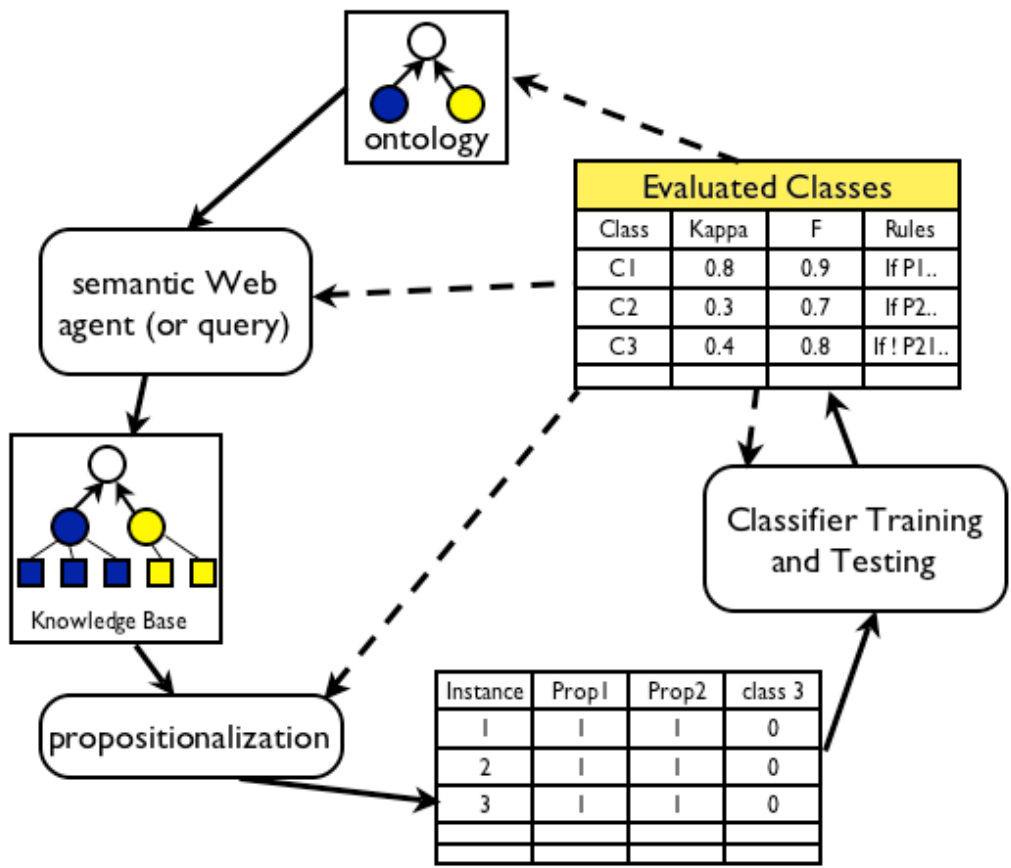

**Figure 3. The complete cycle of ontology evaluation using the OntoLoki method.** The rectangles represent data, the oblong boxes represent processes applied to that data and the arrows represent the order of operations. The dashed arrows indicate how the ontology and any of the contributing processes might be adjusted based on the knowledge gleaned in the evaluation.

### Experiment 1 – control: Phosphabase and its permutations

To evaluate the ability of the OntoLoki method, we began by identifying a gold standard with regard to classification consistency. We sought an ontology and a knowledge base that, in addition to displaying highly consistent classification, were representative of the biomedical domain. A combination that met these criteria was provided by the

Phosphabase ontology [25] and an OWL/RDF knowledge base extracted from UniProt.

Phosphabase is an OWL ontology for describing protein phosphatases based on their domain composition. For example, the class "classical Tyrosine Phosphatase" is defined to be equivalent to the set of proteins that contain at least one InterPro domain IPR000242. Because class definitions like this have been formalized with OWL DL restrictions, it is possible to use reasoners, such as Pellet [9] and Fact++ [8], to automatically classify protein instances within the ontology. For example, if a novel protein is discovered to have an IPR000242 domain, then a reasoner can be used to infer that it is an instance of the class 'classical Tyrosine Phosphatase' [25].

Ontologies, like Phosphabase, that are constructed using DL class restrictions, form a particularly interesting and useful case from the perspective of the quality evaluations suggested here. The formal, computable restrictions on class membership that they encode can be used to guarantee that the instances of each class demonstrate a perfectly specific pattern of properties. The ability to rediscover the decision boundaries formed by these class restrictions and expressed in the properties of instances assigned to the classes can thus serve as a positive control on our implementation. If these straightforward, perfectly consistent patterns cannot be discovered, then the implementation could not expect to identify more complex patterns in noisier, more typical, data.

To create a control knowledge base using a DL ontology like Phosphabase, the following steps are applied:

1. classify the instances of the knowledge base using a DL reasoner
2. strip the definitions from the classes
3. convert the classified knowledge base into a table as discussed above

This process results in a knowledge base where the properties of the instances are guaranteed to contain the information required to identify the decision boundaries represented in the original class definitions but where those definitions are unknown to the implementation.

To prepare the knowledge base for the Phosphabase experiment, instances in the form of proteins annotated with InterPro domains were gathered by a query issued to UniProt beta [18]. The query retrieved reviewed protein records where the term 'phosphatase' appeared in the annotation of the protein in any of the fields: 'domain', 'protein family', 'gene name', 'gene ontology', 'keyword', 'protein name', or 'web resource'. This query, executed on November 6, 2007, produced 3169 protein instances, each with extensive annotation expressed in OWL/RDF.

For the control experiment, the only properties that were considered were the InterPro domains and the presence of transmembrane regions because these were the only features used in the Phosphabase class restrictions. Once the instances were retrieved and their annotations mapped to the Phosphabase representation, they were submitted to the Pellet reasoner and thereby classified within the ontology. Prior to presenting this knowledge base to the evaluation system, the restrictions on class membership were removed from the ontology, but the computed class- assignments for each instance were maintained, thus forming a simple hierarchy of classes with instances consistently assigned according to the (now absent) class definitions. It is worth noting that this final structure mimics that of many existing biomedical ontologies, such as the Gene Ontology, where instances are assigned by curators to a class-hierarchy free of formal definitions.

Table 1 describes the knowledge base constructed for the evaluation of the Phosphabase ontology. The ontology contains a total of 82 classes, of which only 39 define phosphatase classes. The other classes are used to represent things like protein domains and are not included in this evaluation. The root of the classes evaluated was thus the class 'Protein Phosphatase'. Of the 39 phosphatase classes, only 27 had defined restrictions sufficient for inferring class membership and of these only 19 had enough instances to attempt the evaluation. In some cases, the Phosphatase classes used InterPro domains that have now been retired and thus no instances with current annotation could be classified into them. In other cases, there were few instances of that Phosphatase class. Only classes with at least 5 positive and 5 negative examples are evaluated as this is the minimum number required to execute 10-fold cross-validation. In all, only 19 classes (about half of the phosphatase classes) were evaluated.

**Table 1. Phosphabase knowledge base statistics**

| Total classes | 82 |
|---|---|
| Total Protein Phosphatase classes | 39 |
| Total Phosphatase classes with defined necessary and sufficient conditions for class membership (allowing automatic classification) | 27 |
| Total instances | 3169 |
| Total classes with more than 5 positive and 5 negative instances | 19 |
| Fraction Phosphatase classes that could be evaluated based on their instances | 19/39 .49 |

**Phosphabase results**

As expected, the perfectly consistent rules for class membership present for the classes evaluated in this control knowledge base were easily learned by the induction algorithms in nearly every case.

JRip identified rules that accurately reflected the restrictions on class membership in the original ontology for each evaluated class except one. The only class that contained any variance from perfect in the cross-validation runs was the class PPP (phosphoprotein phosphatase). For PPP, JRip learned the rule: "If the protein contains domain IPR004843 or IPR001440, then predict PPP". The original definition for this class corresponded to the rule "If the protein contains domain IPR004843 or IPR006186, then predict PPP". It identified the IPR001440 constraint because the majority of the instances of PPP are actually instances of its subclass, proteinPhosphataseType5 which is defined by a restriction on IPR001440. This results in 1 incorrectly classified instance out of 1462 examples (the instances of PPP's parent Protein_Phosphatase).

Chi25_JRip performed identically to JRip with the exception of the class R2A (Receptor Tyrosine Type 2A). When using all of the data (not in cross-validation), it learned the same rule that covered all of the examples in the database perfectly, "if the protein contains domain IPR0008957 and IPR003599, then predict R2A". However, in one round of 10-fold cross- validation, it learned a different rule and misclassified one of the instances.

OneR, which can only learn one rule using one attribute, worked perfectly on all the classes except PPP and R2A. For PPP, it learned the rule "if IPR004843 then PPP", displaying its inability to add the additional attribute (IPR006186) needed to complete the definition. For R2A, it simply always predicted true in all of the cross-validation runs, thus misclassifying the 12 instances that were R1_R6 phosphatases (R2A's superclass) but not R2As. When applied to all of the training data, it learned the rule "if IPR000884 then R2A" which, though not part of the formal definition of the class, correctly predicted 74 of the 82 instances in the R2A dataset.

Figure 4 shows the average performance of each learning algorithm across each of the 19 classes evaluated in terms of the average accuracy, Kappa statistic, F-measures, and mean Kononenko-Bratko information gain measure as identified in one 10-fold cross-validation experiment. With the exception of the ZeroR algorithm, there is very little difference between the different induction algorithms along any of these metrics.

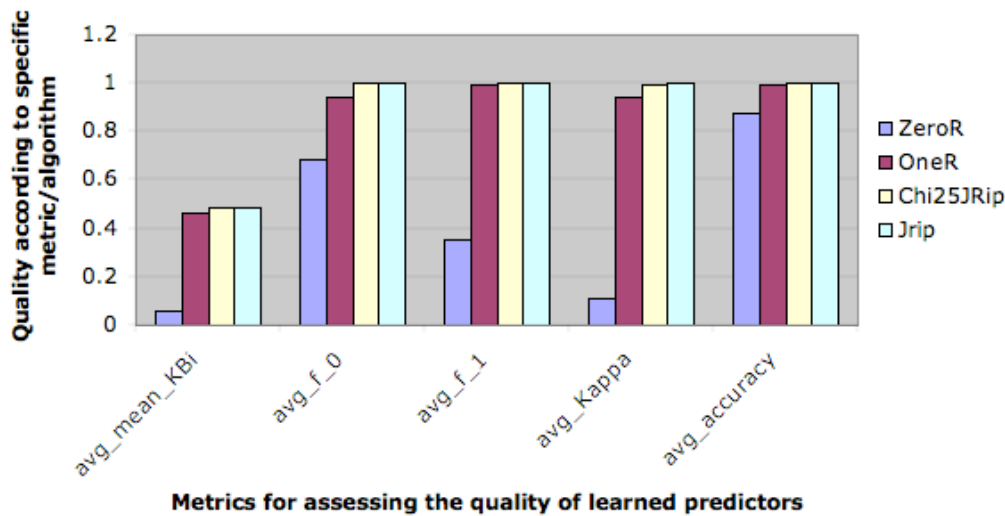

**Figure 4. The average performance of different learning algorithms on predicting the classes in the Phosphabase knowledge base.** Each of the 4 algorithms tested: ZeroR, OneR, Chi25JRip, and JRip, are indicated with a different color. The average performance of the algorithms in 10-fold cross-validation experiments conducted with the Phosphabase knowledge base is shown according to 5 metrics: mean Kononenko-Bratko information gain (KBi), F measure (f_1 measures performance in predicting 'true' for the target class and f_0 indicates performance in predicting 'false'), Kappa, and accuracy.

The ZeroR algorithm doesn't learn from any of the features of the data, it simply makes all predictions for the most frequently observed class in the training set. Hence, it does nothing to expose any defining pattern for the class in question and is thus not useful in terms of assessing the knowledge base. The actual quality of the knowledge base as indicated by any induction algorithm is thus better determined by the difference between its quality and that generated using ZeroR. Metrics that take the prior probability of the test classes into account, such as the Kappa statistic, are thus more effective estimates of the quality of a knowledge base than those that do not.

**Experiment 1 – part 2, permutation-based evaluation**

The next phase of the control experiment was designed to test whether or not the metrics produced using the OntoLoki method are effectively quantifying the quality of the ontology in the context of the selected knowledge base and to provide a preliminary estimate of the expected range of values for the different metrics. Our approach to conducting this test relies on the assumption that a randomly ordered knowledge base (which includes both the ontology and all of the data) should generate lower scores than a non-random knowledge base. Based on this assumption we applied a method similar in nature to that of Brank *et al.*(2006), in which permutations of the assembled Phosphabase knowledge base were generated which contained increasing numbers of random changes to the class assignments for each instance [26]. For each permutation, the amount of noise added was quantified by counting the number of differences between it and the original knowledge base and dividing this number by the total number of instances. If the quality metrics do not correlate with the presence of this added noise, then they clearly do not provide effective measurements of quality.

Figure 5 indicates the relationship between the amount of noise added to the Phosphabase knowledge base and the score for each of the above metrics for the Chi25_JRip algorithm. It shows that each reported metric displays a strong correlation with the amount of randomness added to the ontology, however, the correlation was strongest for average accuracy (r-squared = 0.9994). Based on the trends observed in the chart, the other metrics that seemed to best indicate the presence of the noise in the knowledge base were the Kappa statistic, which consistently varied from a perfect score of 1 all the way down to below 0 (it can range from 1 to –1) and the mean Kononenko-Bratko information-gain measurement which smoothly varied from approximately 0.5 to 0.

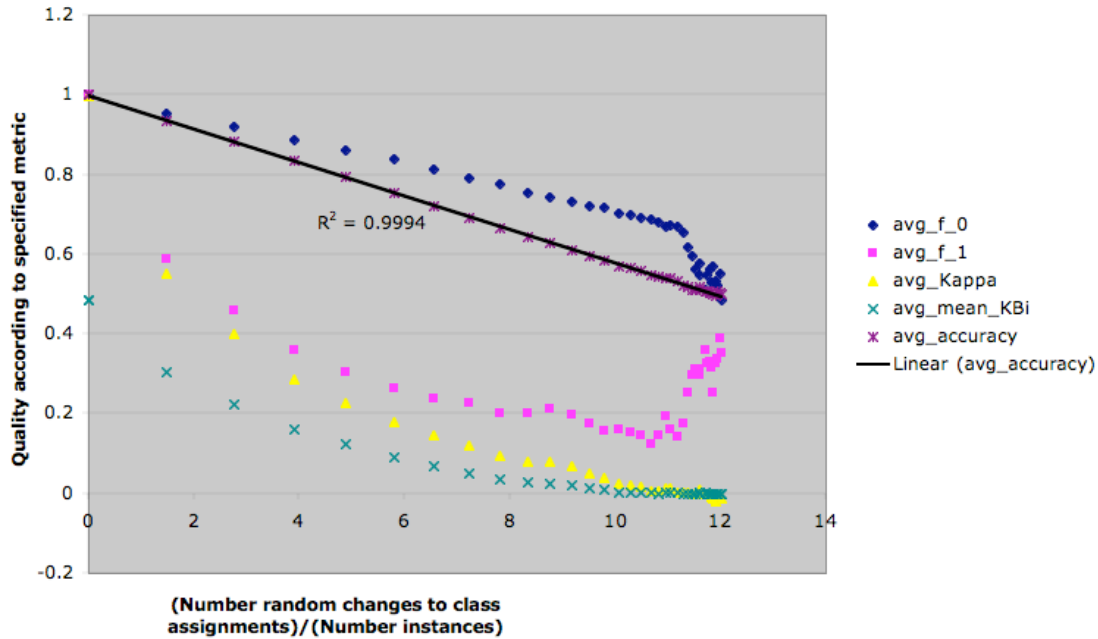

**Figure 5. Results for the Chi25_JRip algorithm as increasing amounts of noise are added to the Phosphabase knowledge base.** Performance is again assessed according to the average F measures, Kappa, KBi, and accuracy.

When the amount of noise added to the ontology reached approximately 11 (which corresponds to about 35,000 random changes), the scores for the f-measure for predicting the class under evaluation ('f_1' on the chart) reversed their expected downwards trend and began to improve. This correlates with the point at which the ability to predict the negative instances for a particular class (instances of its parent that aren't instances of it) begins to decrease rapidly. This somewhat surprising behaviour highlights an influential characteristic of the simple approach to the knowledge base destruction used here and an important weakness of the f-measure in this context.

The simple algorithm to degrade a knowledge base repeatedly a) selects an instance at random, b) selects a class attribute at random, c) checks whether that instance is labelled with that class; if it is, removes the label and if not, adds it. It does not pay attention to the subsumption hierarchy. After running it to saturation, as we did here, the probability for an instance to be labelled with any class in the ontology approaches 0.5 and there is no longer any relationship between classes. So, when creating the local training set for a particular class, the expectation is that 1/2 of the instances in the entire knowledge base will be selected based on the chance of being an instance of that class's superclass and then, of these, 1/2 are likely to be instances of the class being evaluated.

We suggest that the point where F(1) measures begin to increase indicates a threshold beyond which no more information is held in the knowledge base. At this point, algorithms that simply predict true half of the time, which is now near the prior probability in every case, can always get approximately half the predictions correct. Figure 6 shows how the Chi25_JRip algorithm begins to mirror the ZeroR algorithm according to the f-statistics as the noise reaches this threshold. The same behaviour is observed for the OneR and the JRip algorithm without attribute selection.

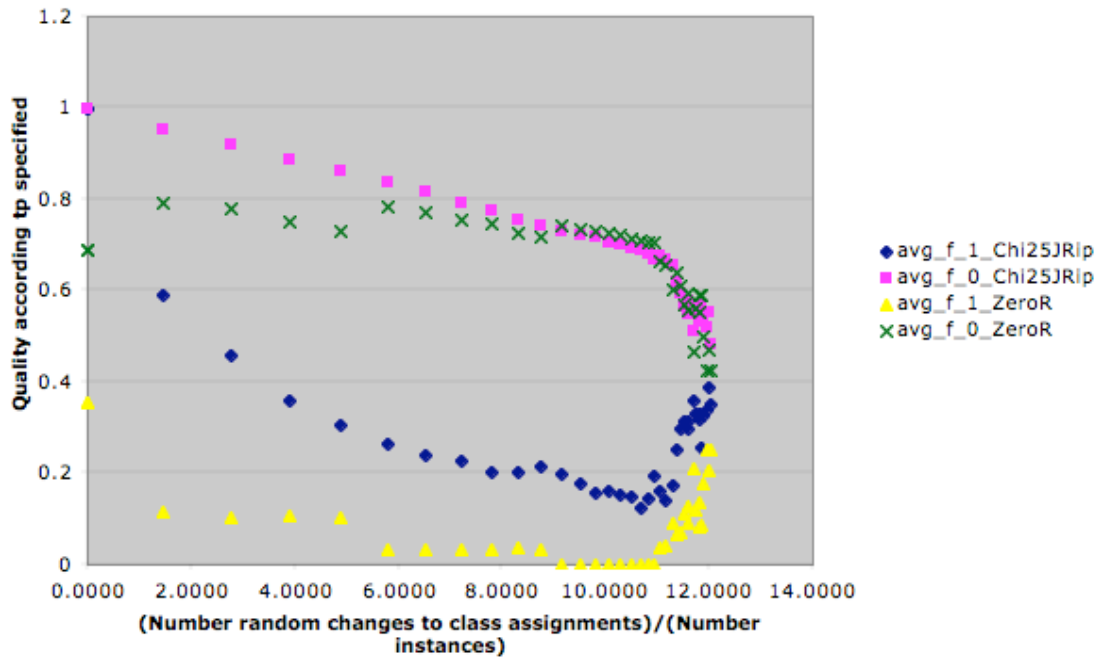

**Figure 6. Effects of increasing noise on the performance of Chi25_Jrip and ZeroR as indicated by the average F measure for each evaluated class.** Note that the results from the sophisticated Chi25_Jrip algorithm eventually begin to mirror those from ZeroR – demonstrating the loss of information from the knowledge base caused by the introduction of random changes.

This behaviour indicates the susceptibility of the F-measures to incorrectly indicating knowledge base quality. According to F(1), one knowledge base that is completely random can receive the same score as one that clearly has some merit as reflected by the other statistics.

**Summary of experiment 1 – control: Phosphabase and its permutations**

The Phosphabase experiments, indicated that:

1. the induction algorithms used in this implementation do effectively learn rules that are predictive of class membership when the information is clearly present in the knowledge base
2. when the required information is removed from the knowledge base, this is reflected in the relative performance of these algorithms
3. the most reliable quality metrics in this context appear to be the Kappa statistic, the Kononenko-Bratko information measure, and the simple percent correct.

In the next section, we assess the implementation's performance in a more typical situation, where the rules for classification are not well known in advance and are not likely to be rigidly defined.

**Evaluating biological ontologies**

In the control experiment we demonstrated that our implementation of the OntoLoki method could successfully find patterns of properties associated with ontology class definitions where they were clearly present. In addition, we showed that the system produced quantitative estimates of classification consistency that correlated with varying levels of knowledge base quality. Given those proofs of concept, we can now begin to explore how the method might be applied to 'real' cases where it is unknown in advance whether the kinds of patterns sought are present in the data. As a beginning in this exploration, we consider two ontologies, one very simple and one fairly complex, that define classifications for the subcellular localizations of gene products. In the first experiment, we conduct an evaluation of an ontology, composed of just six classes in one single level, that represents the classes that can be predicted by the PSORTb algorithm [27]. In the second we move on to the evaluation of the cellular component branch of the Gene Ontology, an ontology composed of more than two thousand classes, arranged in a polyhierachy that in some cases is as much as 11 levels deep.

**Experiment 2 - PSORTb**

PSORTb is a tool for predicting the subcellular localization of bacterial protein sequences [27]. Given a protein sequence, it makes predictions about the protein's likelihood of appearing in one of six locations: the cytoplasm, the cytoplasmic membrane, the outer membrane, the cell wall, the periplasm, or the extracellular space. These classes, which correspond directly to terms from the Gene Ontology, were brought together under the top class 'subcellular localization' to form the very simple ontology used in this experiment.

The knowledge base used to evaluate this ontology, described in Table 2, was created as follows:

1. Instances of proteins and their subcellular localizations were gathered from the hand- curated PSORTb-exp database of experimentally validated subcellular localizations [28].
2. The 'ID Mapping' tool from UniProt beta was used to map the NCBI gi numbers used by the PSORTb database to UniProt identifiers.
3. RDF annotations of these instances were gathered by querying UniProt beta with the UniProt identifiers.

**Table 2. PSORTb knowledge base statistics**

| | |
|---|---|
| Total classes (excluding the top class) | 6 |
| Total instances gathered from PSORTb-exp database | 2171 |
| Total instances used in experiments. Each must have a UniProt identifier and a minimum of at least one annotated InterPro domain or other UniProt annotation property. | 1379 |

The ability of the PSORTb localization prediction system to accurately classify proteins demonstrates that the classes of the generated PSORTb ontology can be predicted for protein instances based on properties associated with those instances. As the algorithms used in PSORTb are different (and much more highly optimized) than the generic rule learners employed here and the features that they process (including the raw protein sequence) are broader than the features used in the current, generic implementation (which does not process the sequences for example), we do not expect to achieve the same high levels of classification performance as the PSORTb implementation. However, there are clearly protein features, such as Signal Peptides, that are captured in the UniProt annotations and are relevant to predicting subcellular localization. From these, it should be possible to identify some predictive patterns automatically and to thus provide this simple, but clearly useful ontology with a fairly good classification consistency score.

The interpretation of what constitutes 'a fairly good score' is still somewhat arbitrary due to the early stage of the development and characterization of the method, but the results indicated by the second phase of the Phosphabase experiment do provide some guidance. Taking the Kappa statistic as an example, Kappas above 0 indicate that the induction process yielded rules that improved upon predictions based only on prior probability and thus the classification could be judged to display at least some degree of property-based consistency. That being said, we expect Kappa scores much higher than 0 in this particular case.

**PSORTb results**

The evaluation identified predictive patterns for each of the classes in the PSORTb ontology. Figure 7 illustrates the average performance of the four learning algorithms tested in cross- validation experiments as described in the previous section. As expected, the JRip algorithm had the best performance across all of the different quality metrics, with the attribute selected version (Chi25_JRip), following closely behind. Figure 8 shows the performance of JRip for each of the classes in PSORTb.

The highest scoring class in terms of simple accuracy of the learned models in the cross-validation runs was 'Cell Wall'; however, this is clearly explained by the relative infrequency of this class within the instances that compose this knowledge base as illustrated by the percentage of positive instances for the class displayed in Figure 8. Since there were only 60 instances of 'Cell Wall' out of a total 1379 in the knowledge base (4%), it is quite easy to achieve a high overall prediction accuracy without learning anything at all by simply always predicting 'not Cell Wall'. This again highlights the importance of accounting for the prior probability of the classes under evaluation when using this system to estimate the quality of a class.

Figure 9 provides a larger view of the performance of JRip using just the Kappa statistic. According to this measure, the best class – in the context of this knowledge base and this induction algorithm – is 'Cytoplasmic Membrane' and the worst is 'Periplasmic'.

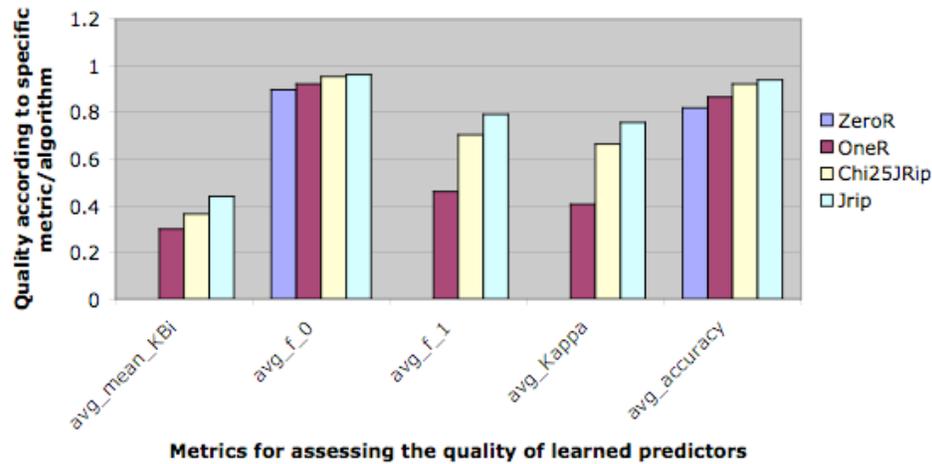

**Figure 7. Average performance of different classifiers according to each reported statistic for the PSORTb classes.** Each of the 4 algorithms tested: ZeroR, OneR, Chi25JRip, and JRip, are indicated with a different color. The average performance of the algorithms in 10-fold cross-validation experiments conducted with the PSORTb knowledge base is shown according to 5 metrics: mean Kononenko-Bratko information gain (KBi), F, Kappa, and accuracy.

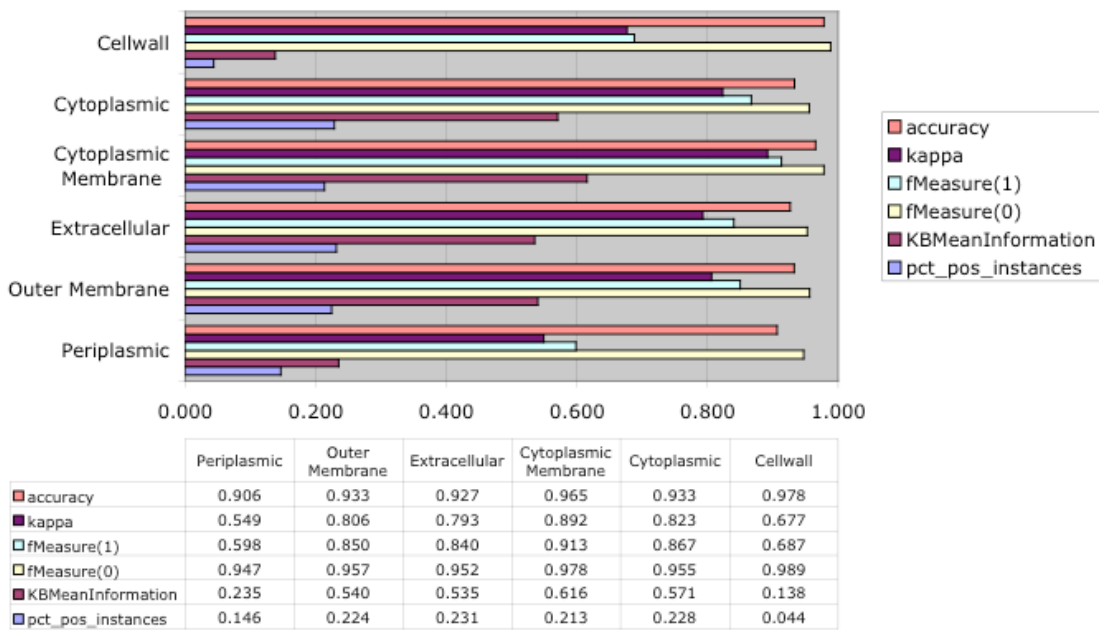

|  | Periplasmic | Outer Membrane | Extracellular | Cytoplasmic Membrane | Cytoplasmic | Cellwall |
|---|---|---|---|---|---|---|
| accuracy | 0.906 | 0.933 | 0.927 | 0.965 | 0.933 | 0.978 |
| kappa | 0.549 | 0.806 | 0.793 | 0.892 | 0.823 | 0.677 |
| fMeasure(1) | 0.598 | 0.850 | 0.840 | 0.913 | 0.867 | 0.687 |
| fMeasure(0) | 0.947 | 0.957 | 0.952 | 0.978 | 0.955 | 0.989 |
| KBMeanInformation | 0.235 | 0.540 | 0.535 | 0.616 | 0.571 | 0.138 |
| pct_pos_instances | 0.146 | 0.224 | 0.231 | 0.213 | 0.228 | 0.044 |

**Figure 8. Classification consistency for PSORTb classes using the JRip algorithm.** Classification consistency is reported for each class using the metrics: accuracy, Kappa, F, and Kononenko-Bratko information gain. In addition, the percentage of positive instances used for each class is presented in blue to give an indication of the baseline frequencies of the different classes in the assembled knowledge base.

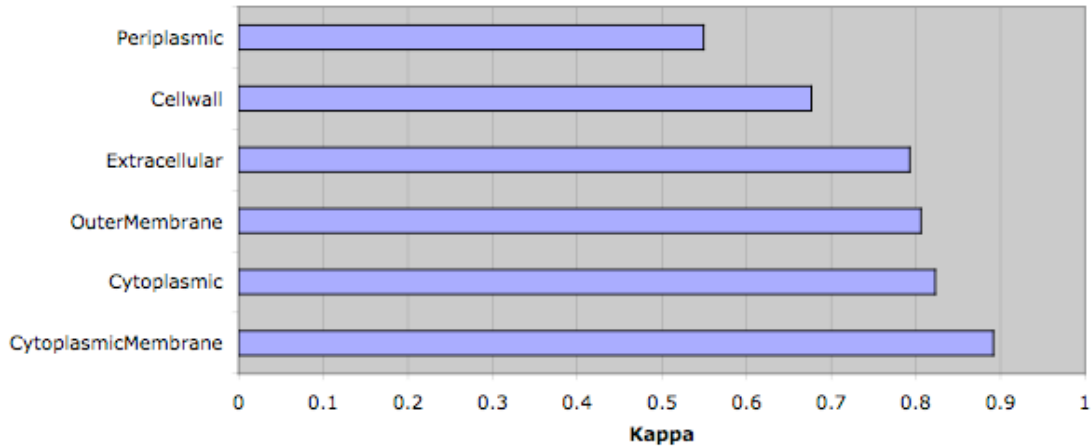

**Figure 9. Inferred classification consistency for PSORTb classes.** The evaluations were conducted using the JRip algorithm and the PSORTb knowledge base. The Kappa statistic was used for the quantification indicated on the X axis.

**PSORTb classification rules learned**

The rules listed below were generated using the JRip algorithm on all the data in the entire training set. We present the complete rule set for 'Cytoplasmic Membrane' and 'Cell Wall' and then samples of the rules covering the most positive instances of the other classes. The two numbers following the consequent of each rule indicate the number of instances classified with the rule and the number of incorrect classifications respectively.

*Cytoplasmic Membrane*

If Transmembrane_Annotation = true, Signal_Peptide_Annotation = false, IPR001343 (Haemolysin-type calcium-binding region) = false and Similarity_Annotation = true Then CytoplasmicMembrane=true (221.0/1.0)

Else if

Transmembrane_Annotation = true, Similarity_Annotation = false and taxon 813 (*Chlamydia trachomatis*) = false Then CytoplasmicMembrane=true (29.0/2.0)

Else if

IPR003439 (ABC transporter-like) = true Then CytoplasmicMembrane=1 (8.0/1.0)

Else

CytoplasmicMembrane= false (1121.0/40.0)

*Cell Wall*

If IPR001899 ('Surface protein from Gram-positive *cocci*, anchor region') = true Then Cellwall=true (31.0/0.0)

Else if

IPR001119 ('S-layer homology region') = true Then Cellwall=true (3.0/0.0)

Else if

taxon = 1423 (*Bacillus subtilis*) and Modification_Annotation = true and Modified_Residue_Annotation = false and part of protein = false Then Cellwall=1 (5.0/0.0)

Else if

IPR010435 ('Peptidase S8A, DUF1034 C-terminal') = true Then Cellwall=true (2.0/0.0)

Else

Cellwall=false (1338.0/19.0)

*Periplasmic*

If Signal_Peptide_Annotation = true and Metal_Binding_Annotation = true and Active_Site_Annotation = false

Then Periplasmic=1 (47.0/7.0)

*OuterMembrane*

If not part of protein and Signal_Peptide_Annotation = true and Similarity_Annotation = true and Modification_Annotation = false and Site_Annotation = false and Subunit_Annotation = true and IPR006059 ('Bacterial extracellular solute-binding, family 1') = false and IPR001782 ('Flagellar P-ring protein') = false

Then OuterMembrane=true (46.0/1.0)

*Extracellular*

Propeptide_Annotation = true and IPR001899 ('Surface protein from Gram-positive *cocci*, anchor region') = false

Then Extracellular=true (110.0/12.0)

*Cytoplasmic*

If Signal_Peptide_Annotation = false and part of protein = true and Transmembrane_Annotation = false and Subunit_Annotation = true and Metal_Binding_Annotation = false

Then Cytoplasmic=true (109.0/2.0)

The first rule learned for predicting localization in the cytoplasmic membrane illustrates the nature of the rest of the rules identified.

1. It describes a large fraction of the instances of cytoplasmic membrane very well, correctly predicting 220 out of the 221 instances that it applies to.
2. It is composed of clearly relevant features such as transmembrane regions and signal peptides, ambiguous features that may be indicators of more specific information not processed by the system such as 'similarity annotation', and features of an uncertain, but potential interesting nature such as IPR001343 (Hemolysin-type calcium-binding region).

**Summary of PSORTb results**

The results of the PSORTb experiment provide evidence that our implementation can identify and quantify the consistency of classification of discriminatory patterns of properties associated with the instances of different classes where those patterns are not known in advance. However, though another useful control on the implementation, the PSORTb ontology is far simpler than the great majority of ontologies in use in the biomedical domain. To characterize our implementation in a realistic situation, we now present an evaluation of a large, well-used, structurally complex, biological ontology.

**Experiment 3 - Cellular Component ontology**

The Cellular Component (CC) branch of the Gene Ontology (GO) provides a controlled vocabulary for the description of the subcellular localizations of gene products [29]. Table 3 provides some basic statistics about the composition of the OWL version used for this evaluation, gathered from the data offered by the experimental Open Biomedical Ontologies (OBO) ontology Web page [30] and the Swoop ontology editor [31]. The version used for the experiment was collected from the OBO Download matrix on Sept. 23, 2007 [32].

The CC ontology was selected for exploratory evaluation for the following reasons.

1. Physical properties, such as protein domains of gene products, are known to be predictive of class membership within this ontology [33]. Thus, in principle, patterns of such properties exist that consistently define membership in at least some of the classes.
2. It is a large and widely used ontology in the biomedical domain with many accessible instances appropriate for the formation of test knowledge bases.
3. It is representative of the structure and basic intent of many other important biomedical ontologies such as several of those being developed as part of the Open Biomedical Ontologies (OBO) initiative [5].

For simplicity, we call the proteins instances of CC classes; however, a gene product is obviously not an instance of a cytoplasmic membrane. Conceptually a gene product may be considered an instance of the class of proteins that tend to localize in the cytoplasmic membrane. These implied classes are what is being evaluated here.

**Table 3. Attributes of the Cellular Component ontology (Sept. 2007)**

| | |
|---|---|
| Total classes | 2127 |
| Average number of direct superclasses per class | 1.49925 |
| Max depth of class tree | 11 |
| Average depth of class tree | 5.9 |
| Max branching factor of class tree (GO_0043234, 'protein complex' has 410 subclasses) | 410 |
| Average branching factor of class tree | 4.4 |

To evaluate the CC, UniProt was once again queried to create an OWL/RDF knowledge base. The query requested protein records that

1. had been annotated with a class from the CC ontology (a subclass of GO_0005575)
2. had been reviewed by UniProt curators
3. had evidence at the protein level (not the fragment level)
4. and had at least one domain annotation

This query resulted in a knowledge base containing 6586 distinct protein instances. Once the knowledge base was assembled, the same evaluation was applied as in the preceding experiments. A specific training/testing table was created for each subclass of the root (GO_0005575) and, assuming it had more then 5 positive and 5 negative instances, was used to train a Chi25_Jrip classifier to attempt to distinguish between the positive and negative instances.

**Cellular Component evaluation results**

In all, only 361 classes (17%) from the CC ontology were evaluated because many classes did not have enough instances in the particular knowledge base extracted from UniProt. The 361 classes had a mean of 1.7 direct (not inferred via subsumption) superclasses. (Recall that, as each class-superclass pair is evaluated independently, per-class scores are averages across each of these evaluations). Within the evaluated classes, a large variability in classification consistency was observed, ranging from classes with perfect defining patterns of properties to classes for which no defining pattern could be identified. The average Kappa score for the 10- fold cross-validation runs across all of the evaluated classes was 0.30, as compared to 0.66 for the PSORTb experiment and 1.0 for the Phosphabase control. Figure 10 provides an illustration of the distribution of the scores observed for the evaluated classes based on the Kappa statistic.

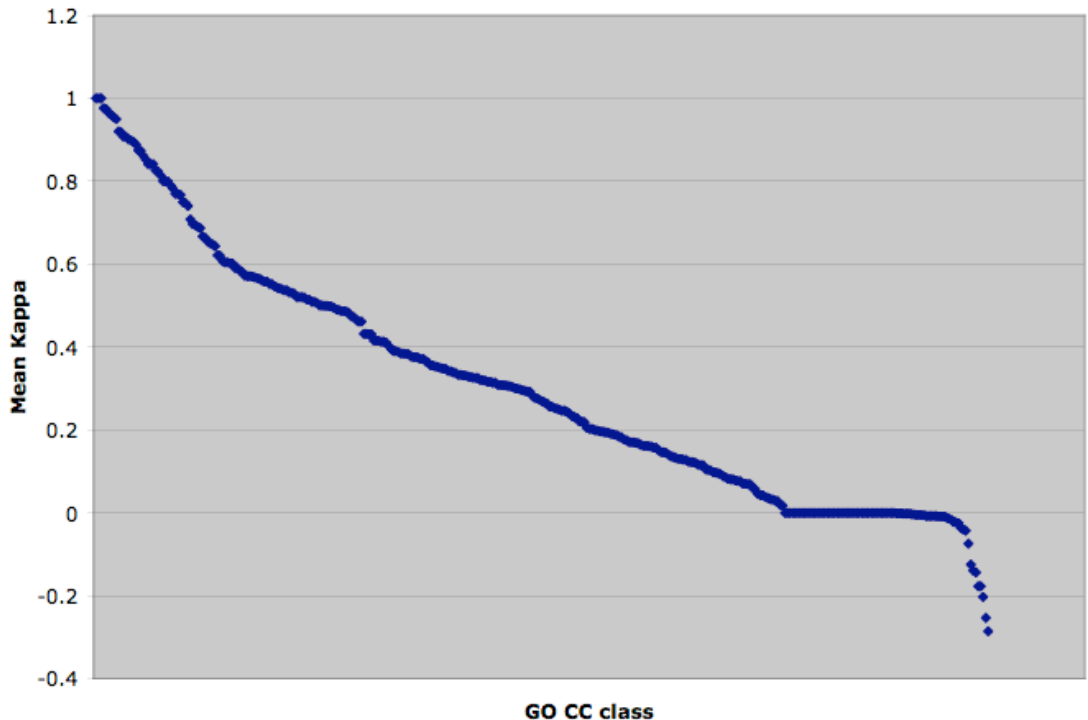

**Figure 10. Classification consistency of a sample of classes from the Cellular Component branch of the Gene Ontology.** The X axis indicates individual classes in the Cellular Component ontology. The Y axis indicates the average Kappa statistic observed for the Chi25_JrRip classifier in 10-fold cross-validation for that class.

**Cellular Component classification rules learned**

The rules learned by the classifiers were often constructed from the taxon and the InterPro domain properties. For example, the first (very simple) rule listed divides the GO class GO_0000142 ('contractile ring (sensu *Saccharomyces*)') from its superclass GO_0005826 ('contractile ring') based on the property 'taxon 4930' (*Saccharomyces*). Below, we list examples of rules identified for some of the highest scoring classes in the cross-validation experiments.

If taxon 4930 (*Saccharomyces*)

       then GO_0000142 ('contractile ring (sensu *Saccharomyces*)') (7/0)

Else GO_0005826 ('contractile ring') (13/0)

If not taxon 3701 ('*Arabidopsis*')

       then GO_0046658 ('anchored to plasma membrane') (10/0)

Else GO_0031225 ('anchored to membrane') (13/0)

If not taxon 4932 (*Saccharomyces cerevisiae*)

    then GO_0005764 ('lysosome') (25/0)

Else GO_0000323 ('lytic vacuole') (17/0)

If IPR001208 ('MCM domain')

    then GO_0042555 ('MCM complex') (24/2)

Else GO_0044454 ('nuclear chromosome part') (92/0)

If IPR002290 (Serine/threonine protein kinase) and IPR015734 ('Calcium/calmodulin-dependent protein kinase 1')

    then GO_0005954 ('calcium- and calmodulin-dependent protein kinase complex')(4/0)

Else if IPR015742 ('Calcium/calmodulin-dependent protein kinase II isoform')

    then GO_0005954 (6/0) Else GO_0043234 ('protein complex') (1157/0)

If IPR013543 ('Calcium/calmodulin dependent protein kinase II, association-domain')

    then GO_0005954 ('calcium- and calmodulin-dependent protein kinase complex')(6/0)

Else if IPR015734 ('Calcium/calmodulin-dependent protein kinase 1')

    then GO_0005954 (5/1)

Else GO_0044424 ('intracellular part') (5071.0/0.0)

If IPR001442 ('Type 4 procollagen, C-terminal repeat')

    then GO_0030935 ('network-forming collagen') (7/0)

Else GO_0005581 ('collagen') (17/0)

If IPR011990 ('Tetratricopeptide-like helical')

    then GO_0031307 ('integral to mitochondrial outer membrane') (5/0)

Else if IPR001806 ('Ras GTPase')

    then GO_0031307 (3/0)

Else GO_00313101 ('integral to organelle membrane') (35/0)

If IPR011990 ('Tetratricopeptide-like helical')

    then GO_0031306 ('intrinsic to mitochondrial outer membrane') (5/0)

Else if IPR002048 ('Calcium-binding EF-hand')

    then GO_0031306 (3/0)

Else GO_0044455 ('mitochondrial membrane part') (21/0)

**Summary of Cellular Component results**

The results of the GO CC evaluation highlight the context-sensitivity of the OntoLoki method. Given a different knowledge base, the results would have been very different. For example, a larger knowledge base would have enabled the evaluation of a much larger proportion of the classes in the ontology. For those classes that were evaluated, the identified rules display a mixture of properties that range from the clearly relevant to the obviously artifactual. Though, upon a very shallow inspection uninformed with any deep knowledge regarding the biology of cellular localization, these results are not overwhelmingly illuminating in terms of understanding or judging the CC ontology, even in their current unrefined form they do provide some insights worthy of consideration. For example, the fact that the taxon properties figured prominently in the delineation of many classification rules indicates the species specific nature of some, but not all, of the definitions of the classes in this ontology. The ongoing work to automatically extract taxon-specific versions of the GO [34], clearly shows that this is an important factor in the evaluation of this ontology. The fact that this basic feature was uncovered automatically by the OntoLoki method thus indicates that the rules discovered can be useful and that their quantification is relevant.

**Discussion**

We presented the OntoLoki method, summarized in Figure 3, for the automatic evaluation of the consistency of classification for ontology classes lacking formal, computable definitions. The implementation begins with an ontology, builds an associated knowledge base, translates that graphically structured knowledge base into a table, and then applies machine learning algorithms to identify specific patterns of properties associated with the different classes. As the dashed lines in Figure 3 indicate, the evaluation of the learned patterns can then be used to make adaptations at each of the steps, starting with the actual ontology and finishing with the algorithms used for induction. Before closing, we highlight some examples of how this method could be applied in several different situations.

**Making use of OntoLoki**

The products of carrying out an OntoLoki evaluation are:

1. a set of rules for assigning instances to the classes in the ontology
2. an estimate of the performance of each inferred classification rule

3. through the rules identified, a direct, empirical assessment of the quality of each class with respect to the implied presence and consistent application of a definition composed of properties present within the knowledge base
4. through the aggregated class scores, an automatic, objective, reproducible quantification of the quality of an ontology in the context of a specific knowledge base

How these products might be applied is highly dependent on the goals of the person conducting the evaluation. We suggest a few ideas, but expect that there are many other possible applications aside from those listed.

The numeric ratings associated with the different classes and whole ontologies could be used for the purposes of,

1. organizing ontology maintenance efforts by identifying classes to attend to based on low classification consistency
2. ordering the results retrieved from ontology search engines [35]
3. evaluating the success of different approaches to the ontology engineering problem

Though the ratings for the classes may be useful, the classification rules likely form the most important product of the system. Such decision models could be used in a variety of different ways, for example,

1. suggesting starting points for the construction of formal (e.g. OWL DL) class definitions within ontologies initially implemented solely as "is a" hierarchies
2. as a means to classify novel instance data within the ontology automatically while leaving the ontology itself unchanged [33].

Aside from improving the ontology or the knowledge base, the knowledge represented in the identified rules could conceivably lead to novel discoveries useful for extending or adapting scientific theory. Scientific ontologies and associated knowledge bases are representations of what is known about the world. By representing knowledge in the form of these 'computational symbolic theories', knowledge can be computed with (tested for internal consistency, integrated, queried) to a much greater extent than would otherwise be possible [36]. The rules found to be associated with class membership thus form both a means for evaluating the classification consistency of an ontology and the opportunity to extend the body of knowledge that it represents.

**Future work**

In the future, it would be useful to improve upon each of the three main phases of the prototype utilized here - the creation of the knowledge base, its propositionalization, and the induction of classification rules from it. The first step, in particular, is crucial. Though the processing of the knowledge base is obviously important, its original creation has by far the most significant effect on the end results of any evaluation. As such, the derivation of efficient methods to dynamically assemble high quality, relevant knowledge bases is a top priority for future investigation. The second step, the propositionalization of the RDF knowledge base, was the most computationally intense aspect of the implementation. In future implementations much more efficient approaches to this step should be identified as the size of the required knowledge bases will only increase. Finally, the induction phase of the method might be improved through the incorporation of learning algorithms specifically designed to infer OWL class definitions from instance data [37] and the integration of techniques from the domain of Formal Concept Analysis that may help to suggest new classes for inclusion in ontologies based on the data used in the evaluation [38].

### 4.3.3 Conclusions

Given the complexity and the diversity of ontologies and their applications, it is unrealistic to expect that a single, universal quality metric for their evaluation will be identified. Rather, as Jeremy Rogers suggests, a comprehensive methodology for ontology evaluation should include assessments along multiple axes [12]. We introduced a new, automated method for the empirical assessment of one aspect of ontology quality that, through its application of automated inductive reasoning, extends and complements existing approaches. Though designed and assessed with the biological and medical domains in mind, the method is applicable in a wide range of other disciplines.